# Electrical control of a Kondo spin screening cloud

Ngoc Han Tu*[1], Donghoon Kim*[2], Minsoo L. Kim*[2], Jeongmin Shim[2], Ryo Ito[1,3], David Pomaranski[4],

Ivan V. Borzenets[5], Arne Ludwig[6], Andreas D. Wieck[6], H.-S. Sim[2],

and Michihisa Yamamoto[1,4]

[1] *Center for Emergent Matter Science (CEMS), RIKEN, Saitama, Japan*

[2]*Department of Physics, Korea Advanced Institute of Science and Technology (KAIST), Daejeon 34141,*

*South Korea*

[3]*National Metrology Institute of Japan (NMIJ), National Institute of Advanced Industrial Science and Technology*

*(AIST), Ibaraki, Japan*

[4]*Quantum-Phase Electronics Center and Department of Applied Physics, The University of Tokyo, Tokyo, Japan*

[5]*Physics & Astronomy Department, Texas A&M University, Texas, United States*

[6]*Faculty of Physics and Astronomy, Ruhr-University Bochum, Bochum, Germany*

*(*) These authors contributed equally.*

Corresponding authors: Ngoc Han Tu (han.tu@riken.jp), H.-S. Sim (hssim@kaist.ac.kr) and Michihisa Yamamoto

(michihisa.yamamoto@riken.jp)

**ABSTRACT.**

Quantitative analysis of quantum many-body systems, consisting of numerous itinerant electrons that interact with localized spins or electrons, is a long-standing issue. The Kondo cloud, a quantum many-body object of conduction electrons that screens a single localized spin, is the building block of such strongly correlated electronic systems. While quantitative analysis of the Kondo cloud associated with a single magnetic impurity is well established for uniform conduction electrons, the fundamental properties of a deformed Kondo cloud influenced by conduction electrons with a modulated density of states remain unsolved. Here we report engineering of the Kondo cloud deformation by confining a part of the cloud into a quantum box called the Kondo box that mimics realistic material systems. We demonstrate quantitative control of the Kondo cloud by developing a way of tuning quantum interference in the box and monitoring the Kondo entanglement. The temperature dependence of the entanglement reveals counterintuitively that the cloud shape is altered mainly outside the box although the quantum interference in the box is tuned. Our work provides a way to simulate various strongly correlated systems by integrating the Kondo cloud, which is not possible in the current theoretical framework.

**INTRODUCTION**

The quantitative analysis of quantum many-body states represents a fundamental challenge. It is particularly true for the strongly correlated electron systems where all numerous itinerant electrons that interact with localized spins or electrons must be treated quantum mechanically. Although strong electronic correlation gives rise to a variety of intriguing phenomena including quantum phase transitions [1-3], spin frustration [4,5], and high-temperature superconductivity [6-8], quantitative analysis of physical quantities of large systems is generally impossible in the present theoretical framework. There is one exception, the Kondo effect for a single magnetic impurity with the constant density states of surrounding conduction electrons, where physical quantities are obtained precisely despite the quantum entanglement of numerous electrons [9-13].

The Kondo effect is the phenomenon of quantum entanglement between a localized impurity spin and conducting electrons. The resulting Kondo cloud is a quantum object extending over the length of several micrometers around the impurity spin and typically contains thousands of conducting electrons when it is realized using a semiconductor quantum dot device [14,15]. The remarkable property of the Kondo cloud is the "universality". Corresponding physical quantities are determined by a single parameter, $T/T_\text{K}$, where $T$ is the temperature and $T_\text{K}$ is the scaling energy of the system called Kondo temperature, irrespective of the detail of the system [16,17]. The spatial extension of the Kondo cloud is also determined by a single parameter, $L/\xi_\text{K}$, where $L$ is the distance from the impurity and $\xi_\text{K} \sim 1/T_\text{K}$ is the size of the cloud, i.e. the cloud has the universal shape [15]. Owing to this universal property, one may expect that various strongly correlated systems can be simulated by designing and integrating the Kondo clouds using semiconductor devices.

In real systems, however, modulation of the density of states brings the Kondo system from the established universal regime to a non-universal regime, where the cloud is deformed, and the physical quantities do not obey the known universal functions of $T/T_\text{K}$. The density of states modulation occurs by various reasons, such as the influence of the localized orbitals, magnetic disorders, and grains, but is mostly understood as the finite size effect. From the opposite perspective, the Kondo cloud can be engineered by controlling such a finite size effect. Operation of a Kondo box [18-25], where the finite box size competes with the Kondo length $\xi_\text{K}$ owing to the modulated density of states in the box, should be a representative way of the Kondo cloud engineering. To watch the result of the engineering, a quantitative determination of the entanglement between the impurity spin and conduction electrons (or the degree of the Kondo screening) using experimentally available observables is desirable. The Kondo box, whose size is smaller than $\xi_K$, has not been realized in experiments yet, due to the lack of a tunable system. It is also known that detection of entanglement is generally a notoriously difficult task. Experimental demonstration of the Kondo box and development of an approach for monitoring the entanglement of the box would be essential steps towards quantum simulation of various correlated electron systems based on tunable Kondo clouds. In what follows, we present experimental and quantitative evaluation of the entanglement in the Kondo box system. The temperature dependence of the entanglement reveals the deformation of the Kondo cloud. Surprisingly, we found that the spatial extension of the Kondo cloud is not so much altered inside the box but is significantly deformed outside the box. The outside extension of the cloud can be engineered via the quantum interference inside the box.

**MODEL**

Our approach to controlling and monitoring the Kondo screening and the entanglement is based on a tunable Kondo system (Fig. 1a). In the system, a quantum dot (QD) hosts an odd number of electrons (whose net spin plays a role of a Kondo impurity spin-1/2 [26-31]) and couples to quasi one-dimensional electron channels. The Kondo box is formed over the distance $L$ between the dot and a quantum point contact (QPC) placed on a channel. We introduce the pinch-off strength $\alpha \in [0,1]$ of the QPC (Eq. (5) in Method-1). $\alpha = 1$ indicates that the box decouples from the rest of the channel, while $\alpha = 0$ means that the box does not form. As the QPC becomes more pinched off (namely, $\alpha$ becomes larger), the electrons become more confined within the box and show a resonance structure of a narrower peak width in their density of states $\rho(E)$. The box is on (resp. off) resonance, when a resonance peak (resp. dip) of $\rho(E)$ is aligned with the Fermi level by tuning $L$. The resonance structure is useful for controlling the Kondo screening, when the box size $L$ is smaller than the bare Kondo cloud length $\xi_{K\infty}$, where the bare cloud length is related to the bare Kondo temperature $T_{K\infty}$ in the absence of the QPC (the $\alpha = 0$ case or $L \to \infty$) as $\xi_{K\infty} = \hbar v_F / k_B T_{K\infty}$, where $k_B$ is the Boltzmann constant, and $v_F$ is the Fermi velocity. In the on-resonance case, the enhanced density of states results in the enhanced Kondo temperature $T_K^{\text{on}}$ ($> T_{K\infty}$), while the Kondo temperature has a reduced value $T_K^{\text{off}}$ ($< T_{K\infty}$) in the off-resonance case.

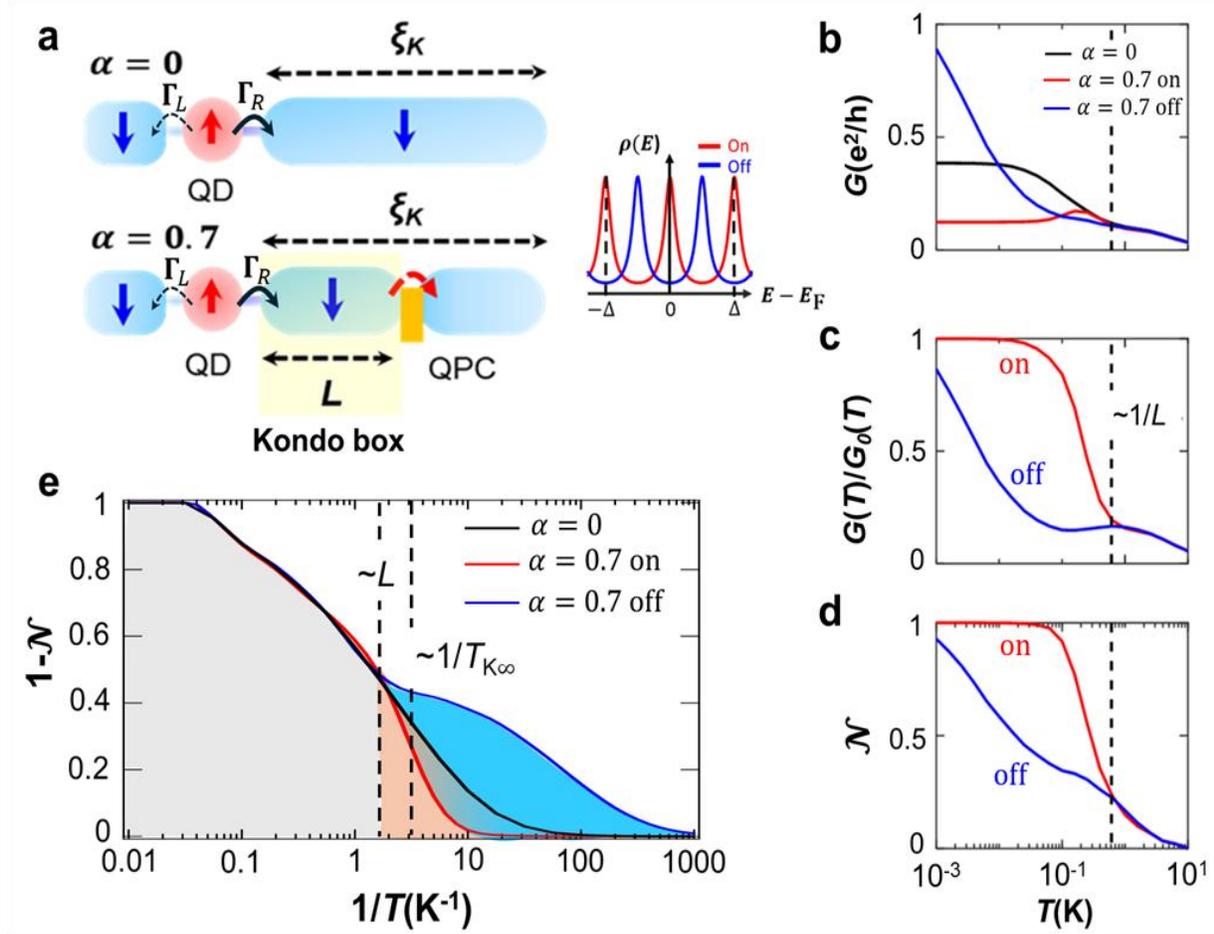

**Figure 1. Kondo box and entanglement determination. a.** Setups in the absence (upper panel) and presence (lower) of a Kondo box. An impurity spin in a quantum dot (QD, red regions) couples with electron channels (blue). The box is formed over the distance $L$ ($< \xi_K$) between the QD and a quantum point contact (QPC) placed on the right channel. $\alpha$ quantifies the QPC strength. In the absence of the QPC, $\alpha = 0$. At larger $\alpha$ ($< 1$), the box is more formed and decoupled from the other part of the right channel. Then the box exhibits on and off resonances in its density of states $\rho(E)$ at the Fermi level $E_F$, depending on its size $L$ (lower panel). $\Delta \sim 1/L$ is the resonance level spacing. **b.** Computation of electron conductance $G$ through the setup as a function of temperature $T$. No box ($\alpha = 0$) and on/off-resonance boxes at $\alpha \sim 0.7$ are considered. **c,d.** The ratio of the Kondo-box conductance $G$ to the reference conductance $G_0$ and the entanglement negativity $\mathcal{N}$ between the QD and the channels, computed for the situations in **b**. **e**. Entanglement negativity $\mathcal{N}$ is plotted as a function of $1/T$, which represents deformation of the Kondo cloud.

Inside the box (a grey area), the spatial extension of the Kondo cloud is not so much altered. Outside the box, the cloud is significantly deformed, depending on the on (a red area) or off (a blue area) situations.

Kondo effects of a QD have usually been identified from the temperature dependence of the electron conductance through the QD. In the case of the Kondo box, the energy-dependent resonance structure of $\rho(E)$ results in the nontrivial temperature dependence $G(T)$ of the conductance through the system at low temperatures. For example, as shown in Fig. 1b, $G(T)$ can have a peak structure in the on-resonance case, which is a combined effect of the Kondo screening, the energy-dependent $\rho(E)$, and asymmetric coupling strengths of the QD to the left and right channels; by contrast, $G(T)$ is monotonic in the off-resonance case. To extract the Kondo screening information from the conductance, we compare the conductance with the conductance $G_0(T)$ of a non-Kondo reference system, which is identical to the Kondo box system except that the QD is replaced by a non-interacting single-particle energy level. We find that the ratio of the Kondo box conductance to the reference conductance is determined only by the Kondo screening, showing the universal Fermi liquid behavior (Supplementary information),

$$G(T)/G_0(T) = 1 - \pi^2 \left(\frac{T}{T_K}\right)^2 + O\left(\left(\frac{T}{T_K}\right)^4\right) \quad (1)$$

at low temperatures of $T \ll T_K^{\text{on,off}}$, where $T_K = T_K^{\text{on (off)}}$ in the on (off) resonance case. The ratio $G(T)/G_0(T)$ depends only on $T/T_K$ and shows the universal thermal suppression of the Kondo screening. On the other hand, the degree of the Kondo screening is quantified by the entanglement between the QD and the rest of the system. We find that the entanglement negativity $\mathcal{N}$ [12,32,33], a measure of the entanglement, also exhibits the universal behavior of $\mathcal{N}(T) = 1 - c\left(\frac{T}{T_K}\right)^2 + O((T/T_K)^4)$ with $c \approx 9.0$ (Supplementary information). Combining the universalities, we propose to determine the Kondo entanglement in experiments by using the conductance ratio $G/G_0$ via the relation,

$$1 - G(T)/G_0(T) = \frac{\pi^2}{c}(1 - \mathcal{N}(T)) + O\left(\left(\frac{T}{T_K}\right)^4\right) \quad (2)$$

at the low temperatures of $T \ll T_K^{\text{on,off}}$. This relation is satisfied, regardless of whether the box is on or off resonance. This is confirmed by our numerical renormalization group (NRG) calculation, as shown in Figs. 1c-1d. Figure S1 shows that the conductance ratio interestingly exhibits qualitatively the same behavior as the entanglement negativity over a wide range of temperature also including $T \gtrsim T_K^{\text{on,off}}$. As $T$ is reduced below the energy spacing $\Delta \sim \frac{1}{L}$ of the box, the conductance ratio and the entanglement increase. This happens more rapidly at higher temperatures for the on-resonance case than the off-resonance, since $T_K^{on}$ is larger than $T_K^{off}$. These justify using the conductance ratio for monitoring the entanglement.

The nontrivial temperature dependence of the entanglement for $T \gtrsim T_K^{on,off}$ is understood as the Kondo cloud deformation as we discuss later. To get intuitive understanding, we plot the entanglement negativity as a function of the inverse temperature in Fig.1e, where the cloud extension grows with the horizontal axis. It shows that the spatial extension of the Kondo cloud remains almost unaltered inside the box (a grey area) while it undergoes a significant deformation outside the box, depending on the on or off situations, shown by the red and blue areas as the shrinkage and extension of the cloud, respectively.

**RESULTS**

**1. Realization of the Kondo box**.

A tunable Kondo system is realized in our device (Fig. 2a). It was fabricated on a high mobility GaAs/AlGaAs heterostructure using the standard Schottky technique (Method-2,3). The conductance $G(T)$ through the device was measured. By tuning the gate voltages $V_P$, $V_R$, and $V_L$ of the QD, the formation of the Kondo state was identified from the behavior that the conductance is enhanced at lower temperatures in a (Kondo) valley region of $V_P$ between two neighboring Coulomb blockade peaks (Fig. 2b and Extended Data Fig. 3). The coupling of the QD to the right (R) 1D channels was made stronger than that to the left (L) channels by tuning $V_R$ and $V_L$. In this regime, the Kondo cloud is expected to be formed mainly in the right channels. We confirmed that the standard Kondo effect appears in our device (with the QPC deactivated) by observing the known temperature dependence of the conductance (Fig. 2d) through the device.

The influence of the QPC placed on the right channels at a distance of $L = 2$ μm away from the QD was then investigated by measuring the conductance at the center of the Kondo valley. Oscillations of the conductance as a function of the QPC gate voltage $V_{QPC}$ were observed at 0.1K (below the bare Kondo temperature $T_{K\infty}$; see below). The oscillation amplitude increases as the QPC is more pinched off with more negative $V_{QPC}$. This implies that the density of states of electrons over the distance $L$ is modulated at the Fermi level as $V_{QPC}$ changes both the distance $L$ and the pinch-off strength $\alpha$. The oscillation dips and peaks are interpreted as alternate on- and off-resonances of the Kondo box, respectively, according to the π/2 phase shift due to the so-called Kondo scattering [15,34,35]. The temperature dependence of the conductance deviates strongly from the case when $V_{QPC}$ is deactivated. This happens when $V_{QPC}$ becomes negative (Figs. 2d-h). From the data of sufficiently small negative $V_{QPC}$ values (Fig. 2e), where the device remains in the standard Kondo regime, we estimated the bare Kondo temperature $T_{K\infty} \sim 0.23$ K, following the approach in Ref. [15] (Method-5). The corresponding Kondo cloud length $\xi_{K\infty} \sim 5.5$ μm is larger than $L$. As $V_{QPC}$ becomes more negative (Figs. 2f-h), the temperature dependence exhibits behaviors more distinct between adjacent $V_{QPC}$ values of dips and peaks in Fig. 2c. As the temperature decreases, the conductance rapidly increases at $V_{QPC} \sim$ -0.3V, -0.5V, -0.6V, while it has lower values at $V_{QPC} \sim$ -0.35V, -0.55 V, -0.65 V and even decreases at $V_{QPC} \sim$ -0.65 V. This behavior is consistent with the off- and on-resonance cases discussed in Fig. 1b, respectively. Indeed, the conductance behavior quantitatively agrees with our NRG calculation of a model whose parameters such as $\alpha$ are selected to fit the calculation results with the measurement data, supporting the Kondo box formation in our device (Method-1). We note that electron-electron interactions within the Kondo box can be ignored in our experiments, as the measured data quantitatively agrees with the NRG calculation in the absence of the interaction.

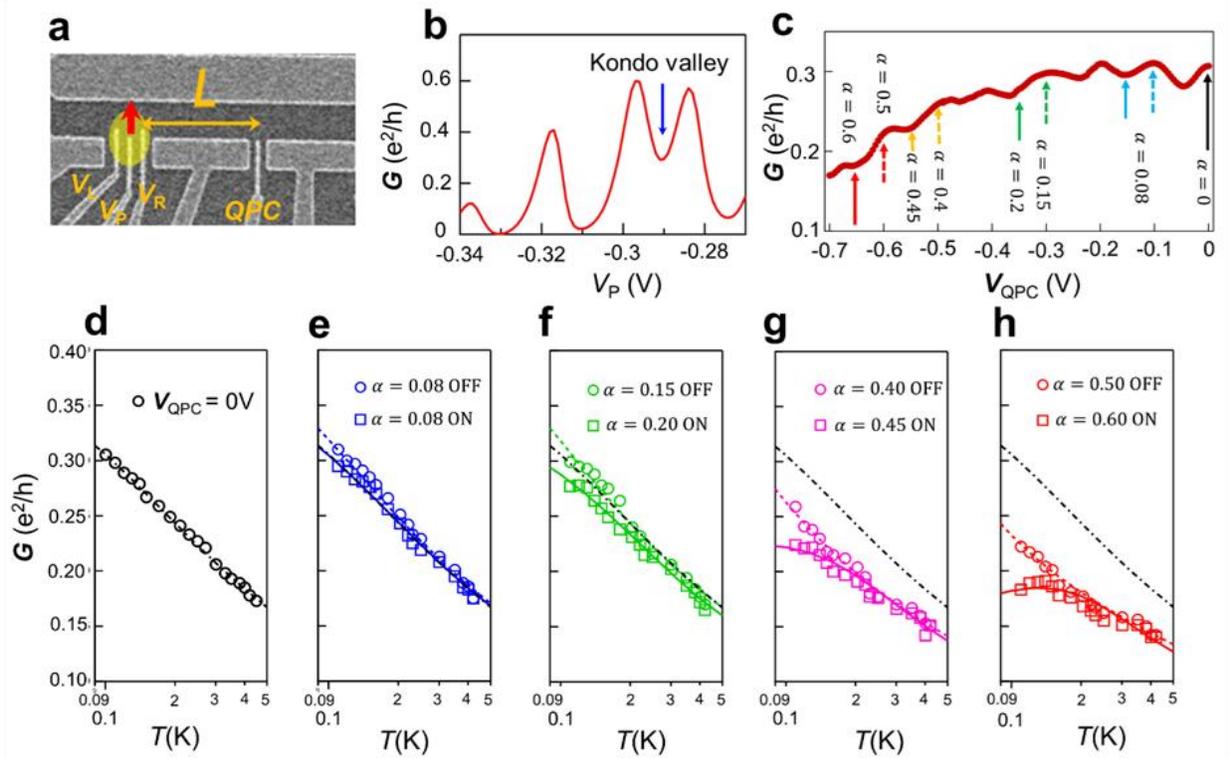

**Figure 2. Realization of a Kondo box. a.** SEM image of the device. A QD is formed by the gate voltages $V_P$ and $V_{R/L}$ while quasi one-dimensional channels are formed by the other side gates that deplete electrons in the gated

regions. A QPC is formed by the gate voltage $V_{\text{QPC}}$ located on the right channels at a distance $L = 2$ μm from the QD. A Kondo box is formed over the distance, when the QPC is activated. **b**. Electron conductance $G$ through the device as a function of gate voltage $V_P$. A Kondo valley is identified. **c**. The conductance $G$ as a function of $V_{\text{QPC}}$ at 0.1K when the QD is in the Kondo valley. **d-h**. Temperature $T$ dependence of the conductance $G$ (open circles and squares) measured at selected values of $V_{\text{QPC}}$ from 0 V **(d)** to -0.65 V **(h)**. The measured data agrees with the theoretical solid (resp. dotted) curves computed for an on-resonance (resp. off-resonance) Kondo box with the values of $\alpha$ marked at the selected values of $V_{\text{QPC}}$ in **c**. The theoretical dashed dot curve for $V_{\text{QPC}} = 0$ V (*i.e.*, $\alpha = 0$) is shown for comparison.

## 2. Monitoring the entanglement and Kondo temperature

To monitor the Kondo screening and the entanglement by using the conductance, we obtain the reference conductance $G_0(T)$ by using the parameters obtained as above (Extended Data Fig. 4). In Fig. 3, we plot the conductance ratio $G/G_0$ and the entanglement $\mathcal{N}$ between the QD and the rest of the device (Method-5 and Supplementary Information). The results show that our experimental data is close to the low-temperature universal regime [36] of $T \ll T_K^{\text{on,off}}$, where $1 - G(T)/G_0(T) = \frac{\pi^2}{c}(1 - \mathcal{N}(T))$ in Eq. (2) is satisfied regardless of whether the device is in the on- or off-resonance situation. In our temperature regime of $T \gtrsim T_K^{\text{on,off}}$, albeit deviating from the universal regime, the two quantities, $1 - G(T)/G_0(T)$ and $\frac{\pi^2}{c}(1 - \mathcal{N}(T))$, show qualitatively the same behavior as a function of the temperature. Hence, it is reasonable to monitor the entanglement by using the conductance ratio. Note that in Fig. 3, the experimental data of $1 - G(T)/G_0(T)$ lie on the theoretical curve over a wider range of $T/T_K$ in the off-resonance situation than in the on-resonance, since the value of $T_K^{\text{off}}$ is more sensitive to $\alpha$ than $T_K^{\text{on}}$ [24].

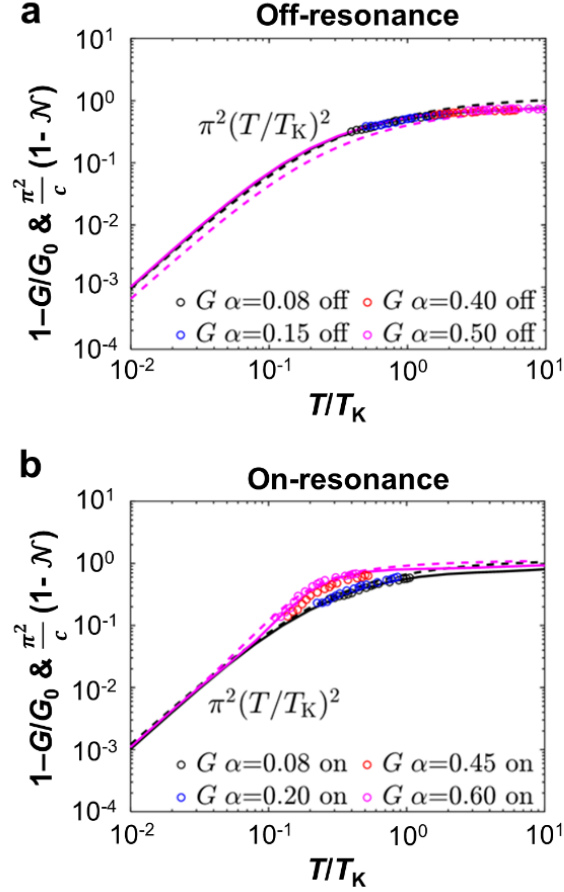

**Figure 3. Monitoring the entanglement from the conductance ratio and extraction of the Kondo temperature.**
**a**. Off-resonance situations. **b**. On-resonance situations. The quantity $1 - G(T)/G_0(T)$ is plotted as a function of temperature, based on the measured values (open circles) and the NRG results (solid curves) at various parameter sets of our device indicated by the values of $\alpha$. It is compared with the values (dashed curves) of the other quantity $\frac{\pi^2}{c}(1 - \mathcal{N}(T))$ with $c \approx 9.0$, obtained from the NRG results of the entanglement negativity $\mathcal{N}(T)$ in the cases of $\alpha = 0.08$ and 0.5 in **a** and $\alpha = 0.08$ and 0.6 in **b**; the curves of $\frac{\pi^2}{c}(1 - \mathcal{N}(T))$ in the other cases of $\alpha$ lie between the two dashed curves in each panel. The comparison shows (i) that the two quantities are equal in the low temperature universal regime of $T \ll T_K$ and (ii) that the experiments are close to the universal regime.

In Figs. 4a and 4b, the conductance ratio $G(T)/G_0(T)$ is compared between adjacent on- and off-resonance situations. Contrary to the conductance $G(T)$, the ratio $G(T)/G_0(T)$ of an on-resonance situation is larger than that of an adjacent off-resonance situation. In the on-resonance situation, the Kondo screening is enhanced with larger Kondo temperature, in comparison with the case of $V_{\text{QPC}}$ deactivated, while in the off-resonance situation, the screening is reduced with smaller Kondo temperature. The enhancement and reduction of the Kondo screening are quantified by the entanglement negativity in Figs. 4c and 4d. The difference of the entanglement (and the Kondo screening) between the on- and off-resonance situations becomes larger as the pinch-off strength $\alpha$ increases. The difference becomes significant at temperatures of $T_K^{\text{off}} < T < T_K^{\text{on}}$; we note that $T_K^{\text{off}} = 0.040$ K at $\alpha = 0.5$ and $T_K^{\text{on}} = 0.980$ K at $\alpha = 0.6$ in Fig. 4d, while the bare Kondo temperature is $T_{K\infty} \sim 0.23$ K. As the temperature decreases below $T_K^{\text{off}}$, the conductance ratio and the entanglement in both the on- and off-resonance situations approach to the maximum values, $G(T)/G_0(T) \to 1$ and $\mathcal{N}(T) \to 1$, following the same universal behavior in Eq. (1), but with different Kondo temperatures. The results in Fig. 4 demonstrate the tunability of the Kondo screening and the entanglement in our device.

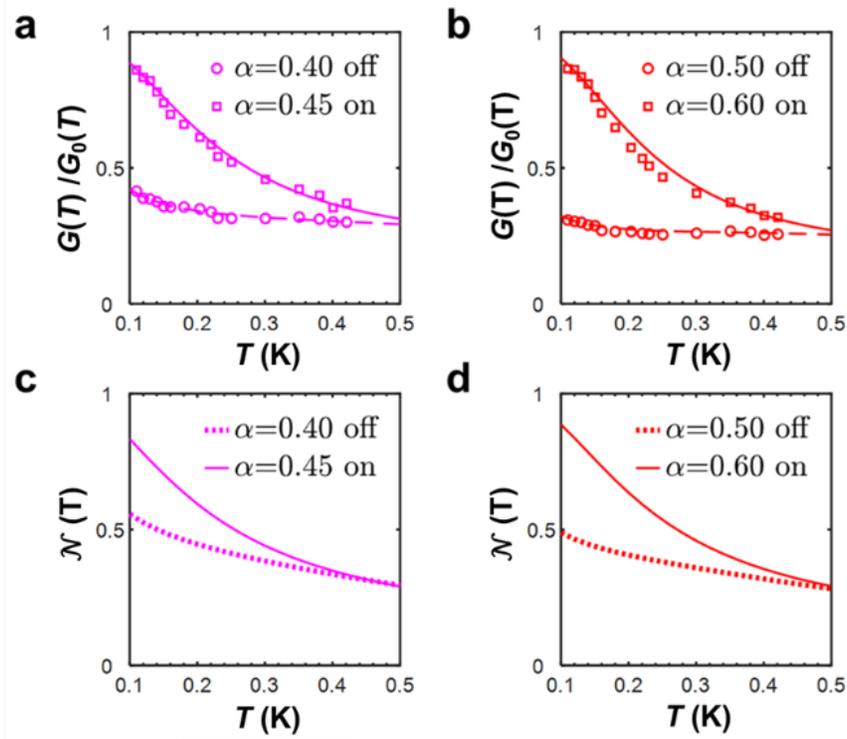

**Figure 4. Controlling the entanglement by changing the QPC. a,b.** Temperature dependence of the conductance ratio $G(T)/G_0(T)$ for two pairs of neighboring on- and off-resonance situations (indicated by the values of $\alpha$) of our device, which are achieved with varying $V_{\text{QPC}}$. The conductance values are from the measurement (open circles and squares) and the NRG calculation (solid and dashed curves). **c,d.** Temperature dependence of the entanglement negativity $\mathcal{N}(T)$, obtained from the NRG calculation in the situations of **a** and **b**.

**DISCUSSION**

We can interpret our findings in terms of the entanglement formation and its spatial extension in our Kondo box of $L < \xi_{K\infty}$ at different temperature regimes (Fig1.e). At high temperature of $T/T_K^{on,off} \to \infty$, $\mathcal{N}(T) \to 0$. As $T$ is reduced, the entanglement is grown from the vicinity of the impurity, with its extension limited by short spin coherence time in thermal scattering processes [12,37]. This trend lasts until $T \sim 1/L$ when the extension approaches the size of the Kondo box. In this temperature regime, the influence of the QPC is not significant. As the temperature becomes lower than $1/L$ and crossing $1/\xi_{K\infty}$, both $G(T)/G_0(T)$ and $\mathcal{N}(T)$ exhibit the non-trivial temperature dependence, implying the formation of the Kondo cloud strongly deformed across the length scale of $\xi_{K\infty}$. While the shape of the Kondo cloud in this regime is not yet understood, it is evident from the rapid (slow) increase of $\mathcal{N}(T)$ that the Kondo cloud of the on-resonance (off-resonance) case is significantly shrunk (prolonged) from that of the universal shape with the length $\xi_{K\infty}$. At low temperature $T \ll T_K^{on,off}$, the entanglement further extends spatially, and $G(T)/G_0(T)$ and $\mathcal{N}(T)$ regain the universal Fermi liquid behavior, but now governed by the energy scale $T_K^{on,off}$. In this low temperature regime, the tail of the Kondo cloud grows with decreasing temperature, implying that the cloud tail outside the Kondo box has a universal shape but with different sizes and intensities between the on-resonance and off-resonance cases.

Our work demonstrates electrical control of Kondo screening at a location 2 μm away from the Kondo impurity. By weakly confining a part of the Kondo cloud within a Kondo box, we could induce a modulation of the density of states of the conduction electrons in a well-controllable manner. We developed an experimental method to quantitatively monitor the entanglement and analyze its temperature dependence. The analyzed data show that while the spatial extent of the Kondo cloud remains unchanged inside the box, it undergoes significant deformation outside the box. Remarkably, this external spatial extension can be precisely tuned - either expanded or shrunk – simply using a local electrostatic gate. Our work represents the first experiment to quantitatively evaluate and control many-body quantum entanglement involving thousands of electrons in a solid-state system. This approach has broad applicability to various many-body systems where coherent coupling between distant localized objects is mediated by conduction electrons through quantum entanglement [38-43]. This paves the way for the quantum simulation of strongly correlated systems, where numerous itinerant electrons interact with localized spins or electrons.

## ACKNOWLEDGMENTS

H.-S. S. is supported by Korea NRF (grant number 2023R1A2C2003430; SRC Center for Quantum Coherence in Condensed Matter, grant number RS-2023-00207732). M. Y. acknowledges CREST-JST (grant number JBMJCR1876). N. H. T., R. I., D. P., and M. Y. acknowledge JSPS KAKENHI (Grant Number JP24H00047).

**Author contributions:** N. H. T. fabricated and characterized the device, performed the experimental measurement, and analyzed the data. D. K., M. L. K., J.S., and H.-S. S. developed the theoretical model. D. K., M. K., and H.-S. S. established the approach to monitoring the entanglement and analyzed the data. R. I., D. P., and I. V. B. contributed to the experimental setup. A. L. and A.D.W. designed and grew the 2DEG wafer. M. Y. and H.-S. S. supervised the project. All authors were involved in discussing results and preparing the manuscript.

## METHOD

### 1. Theoretical model

In this section, we describe the theoretical model for the experimental setup and the details of the numerical renormalization group (NRG) calculation.

*1.1. Model Hamiltonian and local density of states*

The experimental setup consists of the quantum dot (QD), the left and right channels, and the quantum point contact (QPC) at distance $L$ from the QD in the right channel (Extended Data Fig. 1). Its Hamiltonian is written as:

$$H = H_{\text{QD}} + H_{\text{L}} + H_{\text{L-QD}} + H_{\text{R}} + H_{\text{R-QD}} + H_{\text{QPC}} \quad (3)$$

The QD is described by an Anderson impurity [28]. Its Hamiltonian is $H_{\text{QD}} = \sum_\sigma \epsilon_d d_\sigma^\dagger d_\sigma + U n_{d\uparrow} n_{d\downarrow}$, where $d_\sigma^\dagger$ is the operator creating an electron having spin $\sigma$ in the single-particle level $\epsilon_d$ of the QD, $n_{d\sigma} = d_\sigma^\dagger d_\sigma$ is the electron number operator, and $U$ is the Coulomb repulsion energy. Electrons in the left (L) and right (R) channels are described by the semi-infinite tight-binding model. Their Hamiltonian is $H_{\text{L/R}} = -t \sum_\sigma \sum_{i=1}^\infty (c_{\text{L/R},i,\sigma}^\dagger c_{\text{L/R},i+1,\sigma} + c_{\text{L/R},i+1,\sigma}^\dagger c_{\text{L/R},i,\sigma})$, where $c_{\text{L/R},i,\sigma}^\dagger$ creates an electron having spin $\sigma$ in the site $i$ of the left/right channel, and $t$ is the strength of electron hopping between neighboring sites. The channel-dot coupling is described by $H_{\text{L/R-QD}} =$ $t_{L/R} \sum_\sigma (c^\dagger_{L/R,1,\sigma} d_\sigma + d^\dagger_\sigma c_{L/R,1,\sigma})$, where $t_{L/R}$ is the coupling strength between the left/right channel and QD. The QPC is located between the sites $L$ and $L+1$ in the right channel. It is described by $H_{QPC} = -(t_{QPC} - t) \sum_\sigma (c^\dagger_{R,L,\sigma} c_{R,L+1,\sigma} + c^\dagger_{R,L+1,\sigma} c_{R,L,\sigma})$, where $t_{QPC}$ is the electron hopping strength through the QPC.

The box region of the distance $L$ between the QD and the QPC results in a resonance structure of the local density of states (LDOS). The LDOS $\rho_{L/R}(\epsilon)$ at the first site $i = 1$ of the channel L/R (the neighboring site of the QD) are (see, e.g., [15])

$$\rho_L(\epsilon) = \frac{1}{2\pi t}\sqrt{1 - (\tfrac{\epsilon}{2t})^2},$$

$$\rho_R(\epsilon) = \frac{\frac{1}{\pi t}\sqrt{1-(\tfrac{\epsilon}{2t})^2}}{(1-\alpha)\cos^2\left(\tfrac{\pi\epsilon}{\Delta}+k_F L\right)+\frac{1}{1-\alpha}\sin^2\left(\tfrac{\pi\epsilon}{\Delta}+k_F L\right)} \quad (4)$$

where $\Delta = \frac{\pi \hbar v_F}{L}$, $v_F$ is the Fermi velocity, $k_F$ is the Fermi momentum, and

$$\alpha = 1 - (t_{QPC}/t)^2 \quad (5)$$

is the QPC pinch-off strength. When $k_F L = n\pi$, the LDOS shows a resonance peak at the Fermi level (on-resonance). When the $k_F L = (n + \tfrac{1}{2})\pi$, it shows a resonance dip at the Fermi level (off-resonance). Note that we choose the Fermi level $\epsilon_F = 0$. The resonance peak position (or the on/off resonance situation) and the resonance level spacing can be tuned by changing $L$. The resonance width is controlled by changing $t_{QPC}$ (equivalently, $\alpha$).

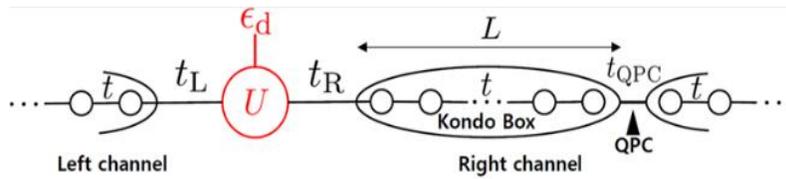

**Extended Data Fig 1. Schematic figure of the Kondo box model.** The QD is represented as an Anderson impurity model with the energy level $\epsilon_d$ and the Coulomb repulsion energy $U$. The QD is coupled to the left and right electron channels, with the coupling strength $t_{L/R}$ for the left/right channel, respectively. The two channels are modeled as tight-binding models with the hopping strength $t$. The QPC is in the right channel, a distance $L$ away from the QD. The presence of the QPC alters the hopping strength of the electron at the QPC position to $t_{QPC}$. Therefore, the Kondo box with size $L$ is constructed in the right channel.

*1.2. Numerical renormalization group calculation*

We calculate the temperature dependence of the conductance $G(T)$ between the left and right channels through the QD, using the numerical renormalization group (NRG) method [44, 45] and based on the relation [46, 47]

$$G(T) = \frac{2e^2}{h} \int d\epsilon \left(-\frac{\partial f}{\partial \epsilon}\right) \frac{4\pi \Gamma_L(\epsilon) \Gamma_R(\epsilon)}{\Gamma_L(\epsilon) + \Gamma_R(\epsilon)} A_{QD} \quad (6)$$

where $\frac{2e^2}{h}$ is the conductance quanta, $f$ is the Fermi-Dirac distribution, $\Gamma_{L/R}(\epsilon) = \pi t_{L/R}^2 \rho_{L/R}(\epsilon)$ is the hybridization function between the QD and the left/right channel, and $A_{QD}$ is the spectral function of the QD.

To compute the $A_{QD}$ by using NRG, we use the full-density matrix formalism [48, 49] and adaptive broadening scheme [50]. Since we consider a low-energy regime, the $\sqrt{1 - (\frac{\epsilon}{2t})^2}$ factor in Eq. (4) is considered as constant. We use the NRG discretization parameter $\Lambda = 4$, the number of kept states $N_{keep} = 500$, and the z-average trick with $z = 0, 0.25, 0.5, 0.75$. We exploit the $U(1) \otimes SU(2)$ charge and spin symmetry to improve the accuracy and the calculation speed by using the QSpace library [51, 52].

We use the parameters $t = 3$ meV, $U = 800$ μeV, and $\Delta = 210$ μeV which are obtained from the experimental data in the regime of $\alpha \ll 1$, following Ref. [15]. When $\alpha \ll 1$, we can obtain $\alpha$ and $t_{L/R}$ from the experimental data by using the method in Ref. [15]. However, the method is only applicable to the $\alpha \ll 1$ regime, we find $\alpha$ and $t_{L/R}$ from the best fit of our computation results of $G(T)$ to the experimental data in the regime of large $\alpha$ (Fig. 2). It is possible to determine $\alpha$ and $t_{L/R}$ from the fit, since the shape of $G(T)$ is affected by change of $\alpha$ and $t_{L/R}$ in a different way. The slope of the curve $G(T)$ is affected by the former, while the overall scale changes with the latter.

Using the parameters found above and the NRG method developed in Ref. [32], we calculate the temperature dependence $\mathcal{N}(T)$ of the entanglement negativity [53] between the QD and the rest of the device.

**2. Experimental set up**

The device was fabricated using a high-quality GaAs/AlGaAs wafer with two-dimensional electron gas located 100 nm below the surface. It has a mobility of $1.47 \times 10^6$ cm$^2$ V$^{-1}$ s$^{-1}$ and an electron density of $n = 1.82 \times 10^{11}$ cm$^{-2}$ at 4K. Experiments were performed in a dilution refrigerator with a base temperature of approximately 50 mK. Electron transport was measured using a standard lock-in technique with a small ac voltage (3-10 μV, $f = 23.3$ Hz) applied to an input contact and a current was measured at an output contact.

In the device, the 1D channels were first formed by applying voltages to the side gates that could fully deplete the carriers under the gates while leaving 1D channels on both sides occupied. Then the QD was formed by tuning three gates: $V_P$, $V_L$, and $V_R$. The plunger gate ($V_P$) modulates the quantum dot's electron population, while the side gates determine the coupling strength to the 1D channels on their respective sides (labelled $V_L$ for the left and $V_R$ for the right). We also slightly tuned a large gate to adjust the size of the QD.

### 3. Electron temperature

The electron temperature $T_{cal}$ is determined from Coulomb blockage measurements [54]. We took a Coulomb peak separated from the Kondo regime and checked the change in the width of this Coulomb peak with increasing the temperature $T_{mix}$ of the mixing chamber. $T_{cal}$ is obtained by fitting the shape of the Coulomb peak width with a function $G \sim \left(\frac{\alpha \Delta V_P}{2 k_B T_{cal}}\right)$. By fitting $T_{mix}$ and $T_{cal}$ with a linear function, the electron temperature $T_{cal}$ is determined (Extended Data Fig. 2).

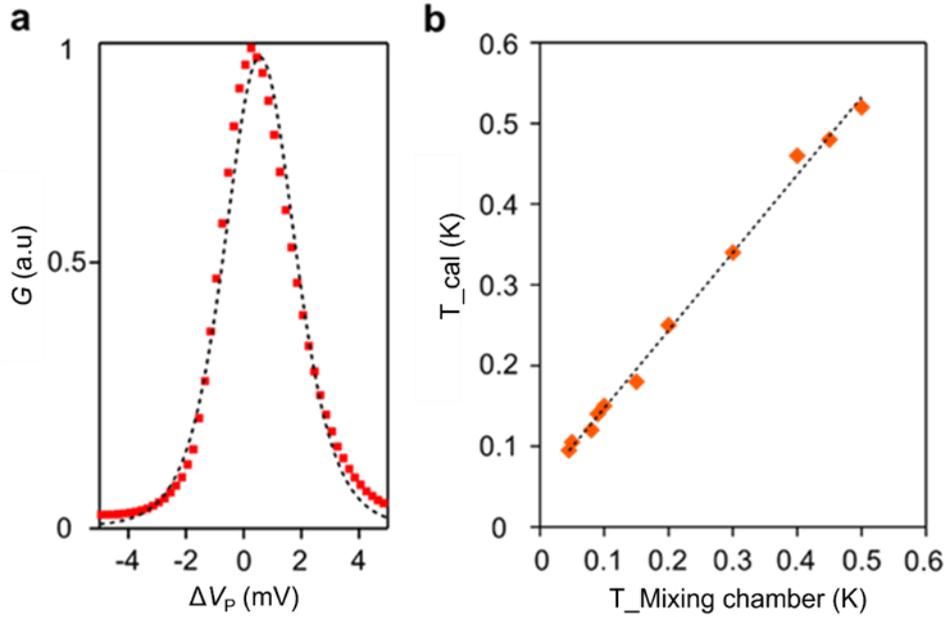

**Extended Data Fig 2**. **Calibration for electron temperature $T_{cal}$. a.** Measured conductance of a Coulomb peak as a function of $V_P$ at the base temperature of the device. We fit the dashed curve to the function of $G \sim \left(\frac{\alpha \Delta V_P}{2 k_B T_{cal}}\right)$, to extract $T_{cal}$. **b.** Relation between $T_{cal}$ and the mixing chamber temperature. It follows a linear fit (dashed line).

### 4. Charging energy

We measure the conductance $G$ across the device as a function of $V_P$ and DC voltage bias $V_{DC}$ in the Kondo regime at the base temperature. Extended Data Fig. 3 shows a Coulomb diamond with a clear maxima conductance between two Coulomb peaks. The charging energy $U \sim 800$ μeV was determined from the height of the Coulomb diamond.

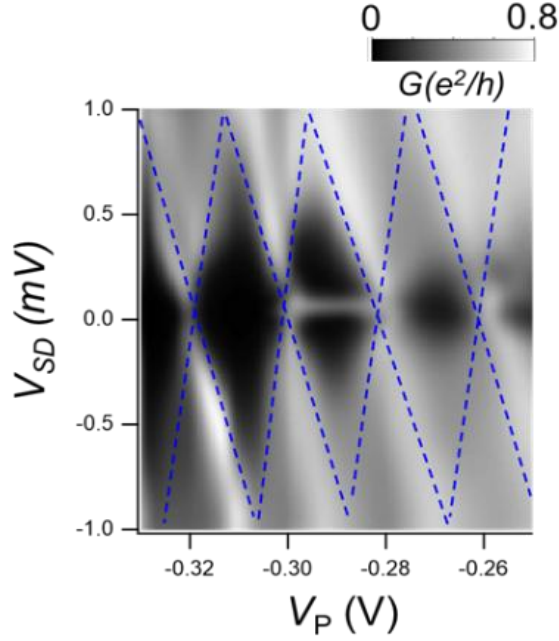

**Extended Data Fig 3.** Determination of the charging energy of the QD from a Coulomb diamond of the conductance across the device. The conductance is plotted as a function of $V_{SD}$ and the plunger gate $V_P$. The charging energy $U \sim 800$ μeV is estimated from the height of the Coulomb diamond.

5. **Calculation of the conductance**

We calculate the temperature $T$ dependence of the conductance $G(T)$ through the QD between the left and right 1D channels by using Eq. (6). While the energy $\epsilon$ dependence of $\Gamma_L$ is ignored for simplicity ($\Gamma_L(\epsilon) = \Gamma_L$), the energy dependence of $\Gamma_R(\epsilon)$ describes the resonance behavior of the Kondo box over the distance $L$.

We obtain the energy dependence of $\Gamma_R(\epsilon)$, based on a model Hamiltonian for the 1D channels and the QPC. We compute $A_{QD}(\epsilon)$ by using the NRG method. The parameters of our model are found by comparing the calculation results with the experimental data. In the regime of $\alpha \ll 1$, $\Gamma_L$, $\Gamma_R(\epsilon)$, and $\alpha$ are extracted from the experimental data by using the approach in Ref. [15]. For example, the Kondo temperatures $T_K^{on}$ and $T_K^{off}$ of the on- and off-resonance situations are obtained by using the empirical formula $G(T) \sim (T_K'^2/(T^2 + T_K'^2))^s$ with $s = 0.22$ [55], and the bare

Kondo temperature is obtained by using $T_{K\infty} = \sqrt{T_K^{on} T_K^{off}}$. Here the relation between $T_K$ and $T_K'$ is found as $T_K = \frac{\pi T_K'}{\sqrt{s}}$, where we adapt the definition of the Kondo temperature $T_K$ from the universal temperature dependence $G(T)/G(T=0) \approx 1 - \pi^2 (T/T_K)^2$ of the conductance in the absence of the Kondo box [56]. Note that this relation is slightly different from the empirical relation $T_K' = T_K / \sqrt{2^{1/s} - 1}$ [55] because of the different definitions of the Kondo temperature. $\Gamma_L$ and $\Gamma_R$ are also estimated from the empirical formula and the Kondo temperature. In the regime of non-negligible $\alpha$, we find $\Gamma_R(\epsilon)$ and $\alpha$ by fitting our NRG results of the conductance to the experimental data.

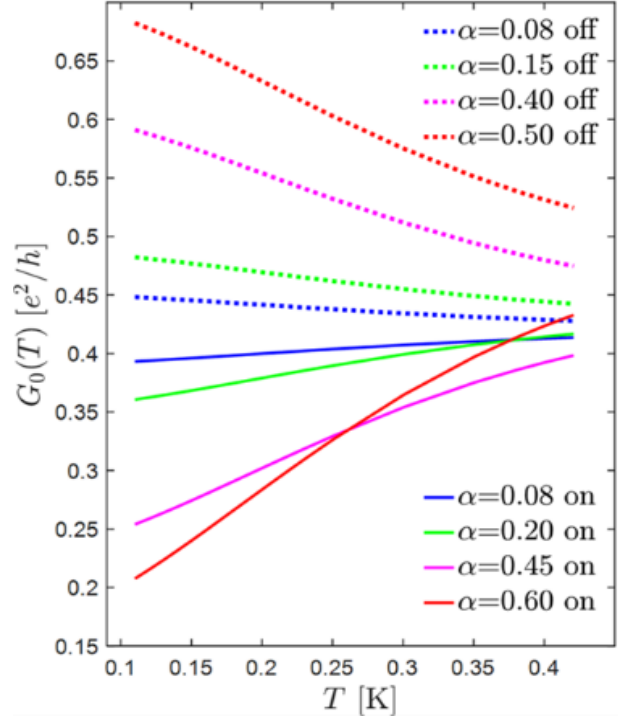

**Extended Data Fig.4 Reference conductance $G_0(T)$.** Its temperature dependence is computed using Eq. (6) and the Breit-Wigner spectral function $A_{BW}(\epsilon, T)$ (Ref. [57]; see Supplementary Information) for the on- and off-resonance situations of various parameter sets (represented by the value of α from 0.08 to 0.6, as shown in the plot). As the temperature decreases, the conductance becomes smaller (resp. larger) in the on-resonance (resp. off-resonance) situations.

**References**

[1] Sachdev, S. Quantum phase transitions. *Phys. World* **12**, 33 (1999).

[2] Vojta, M. Quantum phase transitions. *Rep. Prog. Phys.* **66**, 2069 (2003).

[3] Si, Q. & Steglich, F. Heavy fermions and quantum phase transitions. *Science* **329**, 1161–1166 (2010).

[4] Binder, K. & Young, A. P. Spin glasses: Experimental facts, theoretical concepts, and open questions. *Rev. Mod. Phys.* **58**, 801 (1986).

[5] Balents, L. Spin liquids in frustrated magnets. *Nature* **464**, 199–208 (2010).

[6] Bednorz, J. G. & Müller, K. A. Possible high Tc Tc superconductivity in the Ba-La-Cu-O system. *Z. Phys. B* **64**, 189–193 (1986).

[7] Dagotto, E. Correlated electrons in high-temperature superconductors. *Rev. Mod. Phys.* **66**, 763 (1994).

[8] Capone, M., Fabrizio, M., Castellani, C. & Tosatti, E. Strongly correlated superconductivity. *Science* **296**, 2364–2366 (2002).

[9] Hewson, A. C. *The Kondo Problem to Heavy Fermions* (Cambridge Univ. Press, 1993).

[10] Kondo, J. *The Physics of Dilute Magnetic Alloys* (Cambridge Univ. Press, 2012).

[11] Lee, S.-S. B., Park, J. & Sim, H.-S. Macroscopic quantum entanglement of a Kondo cloud at finite temperature *Phys. Rev. Lett.* **114**, 057203 (2015).

[12] Shim, J., Kim, D. & Sim, H.-S. Hierarchical entanglement shells of multichannel Kondo clouds. *Nat. Commun.* **14**, 3521 (2023).

[13] Yoshida, K. Bound state due to the s-d exchang interaction .*Phys. Rev.* **147**, 223 (1966).

[14] Park, J., Lee, S.-S. B., Oreg, Y. & Sim, H.-S. How to directly measure a Kondo cloud's length. *Phys. Rev. Lett.* **110**, 246603 (2013).

[15] Borzenets, I. V., Shim, J., Chen, J. C. H., Ludwig, A., Wieck, A. D., Tarucha, S., Sim, H.-S. & Yamamoto, M. Observation of the Kondo screening cloud. *Nature* **579**, 210–213 (2020).

[16] Sørensen, E. S. & Affleck, I. Scaling properties of the Kondo screening cloud. *Phys. Rev. B.* **53**, 9153 (1996).

[17] Affleck, I. *Perspectives of Mesoscopic Physics* (World Scientific, 2010).

[18] Thimm, W. B., Kroha, J. & von Delft, J. Kondo Box: A magnetic impurity in an ultrasmall metallic grain. *Phys. Rev. Lett.* **82**, 2143 (1999).

[19] Affleck, I. & Simon, P. D. Detecting the Kondo screening cloud around a quantum dot. *Phys. Rev. Lett.* **86**, 2854 (2001).

[20] Simon, P. & Affleck, I. Finite-size effects in conductance measurements on quantum dots. *Phys. Rev. Lett.* **89**, 206602 (2002).

[21] Simon, P. & Affleck, I. Kondo screening cloud effects in mesoscopic devices. *Phys. Rev. B* **68**, 115304 (2003).

[22] Cornaglia, P. S. & Balseiro, C. A. Transport through quantum dots in mesoscopic circuits. *Phys. Rev. Lett.* **90**, 216801 (2003).

[23] Hand, T., Kroha, J. & Monien, H. Spin correlations and finite-size rffects in the one-dimensional Kondo Box. *Phys. Rev. Lett.* **97**, 136604 (2006).

[24] Simon, P., Salomez, J. & Feinberg, D. Transport spectroscopy of a Kondo quantum dot coupled to a finite size grain. *Phys. Rev. B* **73**, 205325 (2006).

[25] Bomze, Y., Borzenets, I., Mebrahtu, H., Makarovski, A., Baranger, H. U. & Finkelstein, G. Two-stage Kondo effect and Kondo-box level spectroscopy in a carbon nanotube. *Phys. Rev. B* **82**, 161411(R) (2010).

[26] Glazman, L. I. & Raikh, M. E. Resonant Kondo transparency of a barrier with quasilocal impurity states. *JETP Lett.* **47**, 452 (1988).

[27] Ng, T. K. & Lee, P. A. On-site Coulomb repulsion and resonant tunneling. *Phys. Rev. Lett.* **61**, 1768 (1988).

[28] Pustilnik, M. & Glazman, L. I. Kondo effect in quantum dots. *J. Phys. Condens. Matter* **16**, R513 (2004).

[29] Goldhaber-Gordon, D., Shtrikman, H., Mahalu, D., Abusch-Magder, D., Meitav, U. & Kastner, M. A. Kondo effect in a single-electron transistor. *Nature* **391**, 156–159 (1998).

[30] Cronenwett, S. M., Oosterkamp, T. H. & Kouwenhoven, L. P. A tunable Kondo effect in quantum dots. *Science* **281**, 540–544 (1998).

[31] Goldhaber-Gordon, D., Göres, J., Kastner, M. A., Shtrikman, H., Mahalu, D. & Meirav, U. From the Kondo regime to the mixed-valence regime in a quantum dot. *Phys. Rev. Lett.* **81**, 5225 (1998).

[32] Shim, J., Sim, H.-S. & Lee, S.-S. B. Numerical renormalization group method for entanglement negativity at finite temperature, *Phys. Rev. B* **97**, 155123 (2018).

[33] Kim, D., Shim, J. & Sim, H.-S. Universal thermal entanglement of multichannel Kondo effects. *Phys. Rev. Lett.* **127**, 226801 (2021).

[34] Gerland, U., von Delft, J., Costi, T. A. & Oreg, Y. Transmission phase shift of a quantum dot with Kondo Correlations. *Phys. Rev. Lett.* **84**, 3710 (2000).

[35] Takada, S., Bäuerle, C., Yamamoto, M., Watanabe, K., Hermelin, S., Meunier, T., Alex, A., Weichselbaum, A., von Delft, J., Ludwig, A., Wieck, A. D. & Tarucha, S. Transmission phase in the Kondo regime revealed in a two-path interferometer. *Phys. Rev. Lett.* **113**, 126601 (2014).

[36] Yoo, G., Lee, S.-S. B. & Sim, H.-S. Detecting Kondo entanglement by electron conductance. *Phys. Rev. Lett.* **120**, 146801 (2018).

[37] Mitchell, A. K., Becker, M. & Bulla, R. Real-space renormalization group flow in quantum impurity systems: Local moment formation and the Kondo screening. *Phys. Rev. B* **84**, 115120 (2011).

[38] Craig, N. J., Taylor, J. M., Lester, E. A., Marcus, C. M., Hanson, M. P. & Gossard, A. C. Tunable nonlocal spin control in a coupled-quantum dot system. *Science* **304**, 565–567 (2004).

[39] Bork, J., Zhang, Y.-H., Diekhöner, L., Borda, L., Simon, P., Kroha, J., Wahl, P. & Kern, K. A tunable two-impurity Kondo system in atomic point contact. *Nat. Phys.* **7**, 901–906 (2011).

[40] Spinelli, A., Gerrits, M., Toskovic, R., Bryant, B., Ternes, M. & Otte, A. F. Exploring the phase diagram of the two-impurity Kondo problem. *Nat. Commun.* **6**, 10046 (2015).

[41] Iftikhar, Z., Anthore, A., Mitchell, A. K., Parmentier, F. D., Gennser, U., Ouerghi, A., Cavanna, A., Mora, C., Simon, P. & Pierre, F. Tunable quantum criticality and super-ballistic transport in a "charge" Kondo circuit. *Science* **360**, 1315–1320 (2018).

[42] Pouse, W., Peeters, L., Hsueh, C. L., Gennser, U., Cavanna, A., Kastner, M. A., Mitchell, A. K. & Goldhaber-Gordon, D. Quantum simulation of an exotic quantum critical point in a two-site charge Kondo circuit. *Nat. Phys.* **19**, 492–499 (2023).

[43] Piquard, C., Glidic, P., Han, C., Aassime, A., Cavanna, A., Gennser, U., Meir, Y., Sela, E., Anthore, A. & Pierre, F. Observing the universal screening of a Kondo impurity. *Nat. Commun.* **14**, 7263 (2023).

[44] Wilson, K. G. The renormalization group: Critical phenomena and the Kondo problem. *Rev. Mod. Phys.* **47**, 773 (1975).

[45] Bulla, R., Costi, T. A. & Pruschke, T. Numerical renormalization group method for quantum impurity systems. *Rev. Mod. Phys.* **80**, 395 (2008).

[46] Meir, Y. & Wingreen, N. S. Landauer formula for the current through an interacting electron region. *Phys. Rev. Lett.* **68**, 2512 (1992).

[47] Jauho, A.-P., Wingreen, N. S. & Meir, Y. Time-dependent transport in interacting and noninteracting resonant-tunneling systems. *Phys. Rev. B* **50**, 5528 (1994).

[48] Weichselbaum, A. & von Delft, J. Sum-rule conserving spectral functions from the numerical renormalization group. *Phys. Rev. Lett.* **99**, 076402 (2007).

[49] Weichselbaum, A. Tensor networks and the numerical renormalization group. *Phys. Rev. B* **86**, 245124 (2012).

[50] Lee, S.-S. B. & Weichselbaum, A. AAdaptive broadening to improve spectral resolution in the numerical renormalization group. *Phys. Rev. B* **94**, 235127 (2016).

[51] Weichselbaum, A. Non-abelian symmetries in tensor networks: quantum symmetry space approach. *Ann. Phys.* **327**, 2972 (2012).

[52] Zacharias M. and Giustino F. Theory of the special displacement method for electronic calculations at finite temperature. *Phys. Rev. Res.* **2**, 022385 (2020).

[53] Vidal, G. & Werner, R. F. Computable measure of entanglement. *Phys. Rev. A* **65**, 032314 (2002).

[54] Kouwenhoven, L. P., Marcus, C. M., McEuen, P. L., Tarucha, S., Westervelt, R. M. & Wingreen, N. S. Electron transport in quantum dots. In *Mesoscopic Electron Transport, NATO Advanced Study Institutes Series E*, Vol. 345, 105–214 (Kluwer Academic, 1997).

[55] Goldhaber-Gordon, D., Shtrikman, H., Mahalu, D., Abusch-Magder, D., Meirav, U. & Kastner, M. A. From the Kondo regime to the mixed-valence regime in a single-electron transistor. *Phys. Rev. Lett.* **81**, 5225 (1998).

[56] Nozières, P. A. A Fermi-liquid description of the Kondo problem at low temperatures. *J. Low Temp. Phys.* **17**, 31–42 (1974).

[57] Breit, G. & Wigner, E. Capture of Slow Neutrons. *Phys. Rev.* **49**, 519 (1936).